\documentclass[12pt]{iopart}
\usepackage[numbers,sort,comma,compress]{natbib}
\usepackage{iopams}
\usepackage{setspace}
\usepackage{amssymb}
\usepackage{CJK}
\usepackage{graphicx}
\usepackage{subfigure}
\usepackage{multirow}
\usepackage{dcolumn}
\usepackage{enumerate}
\usepackage{makecell}
\usepackage{booktabs}
\usepackage{gensymb}
\usepackage{verbatim}
\usepackage[cal=rsfs,scr=rsfs]{mathalfa}
\usepackage{caption}
\usepackage{dutchcal}
\usepackage{lineno}
\usepackage[colorlinks,linkcolor=blue,anchorcolor=blue,citecolor=blue]{hyperref}
\graphicspath{fig1/}

\begin{document}
\title[Spectrum Estimation for Spectral Mixing: Multi-Ray Model and Weighting Algorithm]{Penumbra-Effect Induced Spectral Mixing in X-ray Computed Tomography: A Multi-Ray Spectrum Estimation Model and Subsampled Weighting Algorithm}

\author{Yifan~Deng\textsuperscript{1,2}, Hao~Zhou\textsuperscript{1,2},and Hewei~Gao\textsuperscript{1,2,*}}

\address{1. Department of Engineering Physics, Tsinghua University, Beijing 100084, China}
\address{2. Key Laboratory of Particle \& Radiation Imaging (Tsinghua University), Ministry of Education, China}
\address{*Author to whom correspondence should be addressed.}
\ead{hwgao@tsinghua.edu.cn}
\vspace{10pt}

\begin{abstract}
	\textbf{Purpose}: With the development of spectral CT, several novel spectral filters have been introduced to modulate the spectra, such as split filters and spectral modulators. However, due to the finite size of the focal spot of X-ray source, these filters cause spectral mixing in the penumbra region. Traditional spectrum estimation methods fail to account for it, resulting in reduced spectral accuracy.
	
	\textbf{Methods}: To address this challenge, we develop a multi-ray spectrum estimation model and propose an Adaptive Subsampled WeIghting of Filter Thickness (A-SWIFT) method. First, we estimate the unfiltered spectrum using traditional methods. Next, we model the final spectra as a weighted summation of spectra attenuated by multiple filters. The weights and equivalent lengths are obtained by X-ray transmission measurements taken with altered spectra using different kVp or flat filters. Finally, the spectra are approximated by using the multi-ray model. To mimic the penumbra effect, we used a spectral modulator (0.2 mm Mo, 0.6 mm Mo) and a split filter (0.07 mm Au, 0.7 mm Sn) in simulations, and used a copper modulator and a molybdenum modulator (0.2 mm, 0.6 mm) in experiments.
	
	\textbf{Results}: Simulation results show that the mean energy bias in the penumbra region decreased from 7.43 keV using the previous SCFM method (Spectral Compensation for Modulator) to 0.72 keV using the A-SWIFT method for the split filter, and from 1.98 keV to 0.61 keV for the spectral modulator. In experiments,  the root mean square error of the selected ROIs was decreased from 77 to 7 Hounsfield units (HU) for the pure water phantom with a molybdenum modulator, and from 85 to 21 HU with a copper modulator.
	
	\textbf{Conclusion}: Based on a multi-ray spectrum estimation model, the A-SWIFT method provides an accurate and robust approach for spectrum estimation in penumbra region of CT systems utilizing spectral filters.
\end{abstract}


\section{Introduction}
X-ray computed tomography (CT) is widely used in modern medical diagnostics and nondestructive testing because it can produce high-resolution, three-dimensional images of an object's interior.
The physical spatial resolution of CT systems, mainly depends on three factors: the pixel size of the detector, the focal spot size of the X-ray source, and the geometry magnification ratio. In particular, the focal spot size influences the resolution through penumbra effects. A larger focal spot size results in blurring at the edges of the scanned object and reduces the system's spatial resolution. Physically reducing the focal spot size presents technical challenges, such as higher focal temperatures and decreased X-ray intensity. 
To address this challenge, researchers have developed various software methods to manage the penumbra effects, such as modeling the finite focal spot\cite{Kueh2016ModelPenumbra,van2020ModelPenumbra,Hu2022ModelPenumbra}, using blind deconvolution\cite{Hehn2019Blind}, and applying deep-learning methods\cite{Yu2024PenumbraDeep,Liu2024PenumbraDeep}. These methods construct focal spot models or blurring kernels, using deconvolution techniques or similar networks to recover high-resolution images.

\begin{figure*}[h]
	\centering
	\includegraphics[width=16cm]{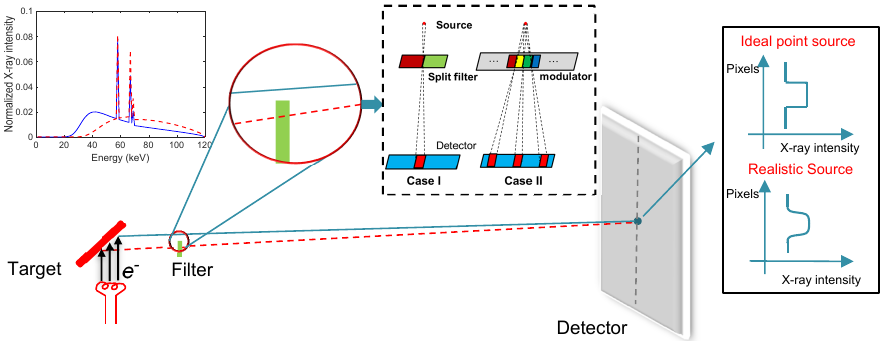}
	\caption{An illustrative diagram of penumbra effects for spectral filters.} \label{fig:penumbra_effect_diagram}	
\end{figure*}

However, in addition to the penumbra effects on spatial resolution, for CT systems with spectral filters, the penumbra effects can also affect the spectra at the edges of the filtration. As Figure \ref{fig:penumbra_effect_diagram} shows, some of the X-rays emitted from the focal spot of the X-ray source may pass through filter 1, while the other part through air (or filter 2), and are both received by the same detector pixel, resulting in spectral mixing. The mixed spectra in the penumbra region complicate the spectrum estimation and negatively affect the accuracy of beam hardening correction, and dual-energy material decomposition.

The spectral mixing induced by the penumbra effect has often been overlooked because the X-ray attenuation of most filters varies slowly, or the edges of the filters are outside the scanning field of view. However, with the advancement of spectral CT and filter-based technologies, filters now come in various forms beyond the standard flat design, including bowtie filters, dynamic filters\cite{Lin2022Dynamic}, and many other types specialized for spectral imaging, such as split filters\cite{Euler2016SplitFilter}, spatial-spectral modulators \cite{Tivnan2021Modulator}, fine grids with varying X-ray incidence angle \cite{Stayman2021modulator}, rotatable filters \cite{Jiang2021Filter},  spectral filtration of a dual-focus CNT X-ray source \cite{Li2022splitfilterCNT}, spectral modulators with flying focal spot \cite{Deng2024Modulator}. These low-cost and portable spectral filters can facilitate spectral imaging in many traditional CT systems, but their X-ray attenuation also exhibit spatial discontinuity, leading to obvious spectral mixing. Figure \ref{fig:penumbra_effect_diagram} also illustrates the penumbra effects of the split filter and modulator. Therefore, with the introduction of more and more spectral filters that exhibit rapid spatial variations in X-ray attenuation, spectral mixing can no longer be ignored. 

Moreover, while both issues arise from the penumbra effect, spectral mixing is a more complex problem than spatial resolution loss. This complexity arises because spectral mixing requires the deconvolution of multiple spectral curves, rather than the deconvolution of multiple projection values. Consequently, traditional deblurring methods used for restoring resolution cannot be directly applied to spectral mixing.
In this case, it is better to directly model the blurring spectra rather than restore the initial spectra. 

A lot of spectrum estimation methods have been proposed for X-ray CT, which can be mainly divided into two categories: model-based and measurement-based methods. The model-based methods\cite{Tucker1991SpecModel,poludniowski2009spekcalc,Omar2020MP,FitzGerald2021MP} directly generate spectra based on physical models, but hard to model the actual non-ideal factors, such as the non-uniformity of the filters and the non-uniformity of detector response; The measurement-based methods\cite{Sidky2005SpecEst,Duan2011SpecEst,Zhao2015PMB,Gao2019SCFM,Pan2023SpecEst} usually employ a step or a wedge phantom to measure data with different thicknesses of known materials, and reconstruct spectra by solving the linear equations that represent the attenuation processes of x-ray photons, which can calibrate the spectral error in model-based methods. However, such a linear system can be ill-positioned as each energy bin of the spectrum is described as an unknown variable, leading to an underdetermined system of equations. In recent years, some indirect measurement-based methods have also been proposed\cite{Chang2020TMI,ChangIEEE2024,Ren2024SpecEst,Lv2024PMB}. These methods take advantage of the reconstructed CT images and the known phantom information as guidance to optimize the spectrum, which provides extra information and improves the solvability of the equations. An improved parameter spectrum model was also proposed recently to reduce the number of variables of the spectrum\cite{ChangIEEE2024}. 

However, for spectral filters with more rapid attenuation changes, these traditional model-based methods are difficult to model the non-uniformity of the filters. Meanwhile, measurement-based methods face challenges in obtaining precise initial spectra, and often struggle to converge to the true spectra as the model of spectral filters are missing. Specifically, for a novel spectral filter known as the spectral modulator, we previously proposed a method called Spectral Compensation for Modulator (SCFM)\cite{Gao2019SCFM}, which uses an initial spectrum with attenuation of single filter for each detector pixel to model the final spectrum. It performs well for spectral modulator with relatively slow variation in X-ray attenuation, as the penumbra effect is relatively mild in such cases. But for modulators with more rapid attenuation changes, this method also contains an inherent error because it neglects the spectral mixing.

To overcome the challenge of spectral mixing, we develop a multi-ray spectrum estimation model that incorporates the penumbra effect, and propose an Adaptive Subsampled WeIghting of Filter Thickness (A-SWIFT) method, which regards the mixed spectra as a weighted summation of the spectra attenuated by multiple filters.

\section{Method}
\subsection{Traditional Model for Spectra Mixing by Filters} 
\label{sec:analysis}
In CT systems, when scatter signals can be ignored, the polychromatic projection $p$ can be modeled using the equivalent spectrum $S(E)$ as,
\begin{eqnarray}
	p = -{\rm ln}\left(\frac{\int S(E){\rm exp}(-\sum_i\mu_i L_i ) ~ dE}{\int S(E)~ dE} \right)
	\label{eq:projection_model}
\end{eqnarray}
where $\mu_i$ is the X-ray attenuation coefficient of the $i_{\rm th}$ material of the scanned object, and $L_i$ is the corresponding equivalent length. With the equivalent spectrum $S(E)$ accurately estimated, we can derive the equivalent length of single material like water by beam hardening correction (BHC), or equivalent lengths of two or more materials by material decomposition. 

For CT systems with filters, the spectrum can be further modeled as a  traditional spectrum estimation method (SCFM\cite{Gao2019SCFM}) described,
\begin{eqnarray}
	S(E) = S_{0}(E){\rm  exp}(-\mu_{\rm F}(E)T_{\rm F})
	\label{eq:SCFM_model}
\end{eqnarray}
where $S_{0}(E)$ is the unfiltered spectrum, $\mu_{F}$ is the X-ray attenuation coefficient of the filter's material, and $T_{F}$ is the corresponding equivalent length, which can be derived from the measured transmission of the filter $A_F$ as,
\begin{eqnarray}
	A_F(T_F) = \frac{\int S_{0}(E){\rm exp}(-\mu_{F}(E)T_{F}) dE}
	{\int S_{0}(E)dE}
	\label{eq:constraint}
\end{eqnarray}

\begin{figure}[h]
	\centering
	\includegraphics[width=16cm]{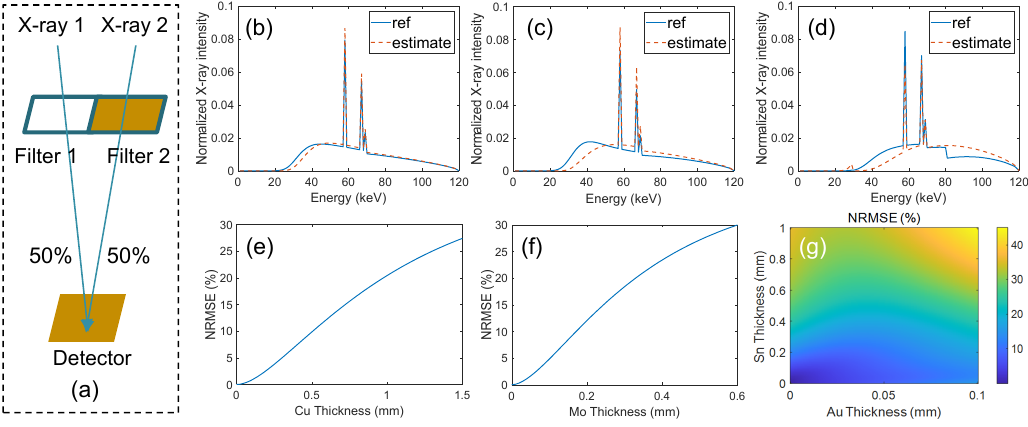}
	\caption{Normalized root mean squared error (NRMSE) of spectra $\widetilde{S_{\rm mix}}$ estimated by SCFM method in a spectral mixing case (reference spectrum: $S_{\rm mix}$). 
	(a) Diagram of spectral mixing in a simplified case; 
	(b) $S_{\rm mix}$ and $\widetilde{S_{\rm mix}}$ in the case of (0, 0.6 mm Cu) mixing; 
	(c) $S_{\rm mix}$ and $\widetilde{S_{\rm mix}}$ in the case of (0, 0.6 mm Mo) mixing; 
	(d) $S_{\rm mix}$ and $\widetilde{S_{\rm mix}}$ in the case of (0.7 mm Sn, 0.07 mm Au) mixing; 
	(e) NRMSE versus copper thickness in the case of (air, Cu); 
	(f) NRMSE versus molybdenum thickness in the case of (air, Mo); 
	(g) NRMSE versus tin and gold thickness in the case of (Sn, Au).} \label{fig:thickness_influence}	
\end{figure}

However, such a model ignores the spectral mixing caused by penumbra effect and results in spectrum estimation error. Here we can analyze the error in a simplified case.  Since the spectrum at penumbra region is a mixture of the spectra of X-rays passing through different side of the edge of the spectral filters, assume the mixed spectrum $S_{\rm mix}$ is the average of the spectrum attenuated by filter 1 $S_{\rm F1}$ and filter 2 $S_{\rm F2}$ as Figure \ref{fig:thickness_influence} (a) shows.

Then, we use equation (\ref{eq:SCFM_model}) to calculate the equivalent length $\widetilde{T_{F}}$ and the estimated spectrum $\widetilde{S_{\rm F}}(E)$. Taking a 120 kVp spectrum as the initial spectrum $S_{0}$, Figure \ref{fig:thickness_influence} (b)-(d) shows the comparison of the reference spectrum and the estimated spectrum of different filters. As normalized root mean square error (NRMSE) is a widely used metric for assessing spectrum estimation accuracy\cite{Chang2020TMI}, we use NRMSE between the actual spectrum $S_{\rm mix}$ and the estimated spectrum $\widetilde{S_{\rm mix}}$ to quantitatively analyze the spectrum estimation accuracy. Figure \ref{fig:thickness_influence} (e)-(g) shows how NRMSE changes with filter thicknesses. From the analysis, the NRMSE can reach nearly 30\% for modulator with blockers like (air, 0.6 mm Mo), and also be up to 30\% for a regular split filter of (0.07 mm Au, 0.7 mm Sn). Such large errors in spectrum estimation result in obvious incorrect CT values. Therefore, there is an urgent need for a new model for these spectral filters.

\subsection{Adaptive Subsampled WeIghting of Filter Thickness (A-SWIFT)}
Considering the influence of the spectral filter, and the difficulty to obtain the precise shape of the focal spot and the exact distribution of X-ray intensity within the focal spot in the original model for precise modeling. We use a weighted multi-ray model instead to approximate the spectrum. Different from the SCFM method, the mixed spectra are modeled as a subsampled weighting of two or more spectra attenuated by multiple filters as,
\begin{eqnarray}
	&S(E) = \sum_{k=1}^{K}w_k \cdot \left(S_0(E) {\rm e}^{-\mu_{\rm k}(E)T_{k}} \right) 
	\label{eq:spec_model_new}
\end{eqnarray}
with,
\begin{eqnarray}
	\sum_{k=1}^{K}w_k=1, w_k \in [0,1]
\end{eqnarray}
Here, subscript $k$ represents the $k_{\rm th}$ material; $w_k$ is the weight coefficient; $T_{k}$ is the corresponding equivalent path-length.

It should be noted that the number of weights $K$, can be more than the number of materials of the spectral filter. For example, in this work we used $K=4$ for the tin-gold split filter, with $\mu_1,\mu_2$ represent the X-ray attenuation of tin, and $\mu_3,\mu_4$ represent the X-ray attenuation of gold. 
Taking advantage of the high degrees of freedom provided by variable weights and thicknesses, this model is able to effectively fit the mixed spectrum. 

Next, we calculate the weights and thicknesses for each detector pixel by using X-ray transmission measurements. 
Without scanning objects, we can employ a step phantom or a series of flat filters with known X-ray attenuation and thicknesses to alter the initial spectrum, or directly use different tube voltages. With the $m_{\rm th}$ altered initial spectrum $S_m(E)$, the measured transmission of the spectral filter $A_m(w,T)$ is given by, 
\begin{eqnarray}
	&A_m(w,T) = \int \sum_{k=1}^{K} w_k \cdot \left(S_m(E) {\rm e}^{-\mu_{k}(E_n)T_{k}} \right) dE 
	\label{eq:transmission_model}
\end{eqnarray}

\subsection{Algorithm Implementation} 
\label{sec:model1} 
\begin{figure}[htb]
	\centering
	\includegraphics[width=16cm] {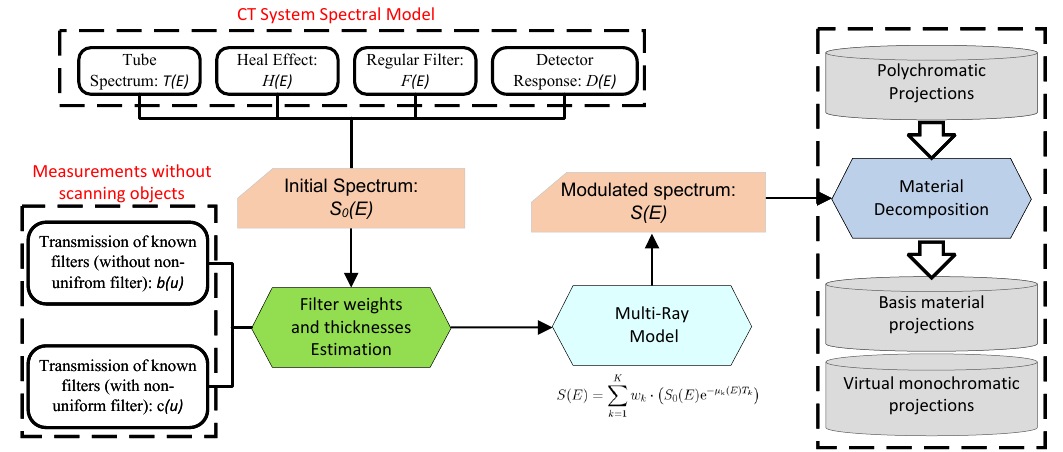}
	\caption{Flowchart of the multi-ray modeling method A-SWIFT.} \label{fig:Flowchart}	
\end{figure}

Based on the multi-ray model, the implementation of the A-SWIFT method can be divided into four steps as shown in Figure \ref{fig:Flowchart}.
\subsubsection{System Spectrum Calculation}
Following the previous analysis on spectral model\cite{Tucker1991SpecModel,Gao2019SCFM}, the spectrum without the spectral filters can be described as,
\begin{eqnarray}
		S_0(E) &= \Gamma(E)\times H(E)\times F(E) \times D(E)
		\label{eq:spec_model}
\end{eqnarray}
where, $\Gamma(E)$ is the X-ray tube spectrum of X-rays emitted from X-ray source, which can be modeled by empirical spectral model methods\cite{Tucker1991SpecModel,Tucker1991SpecModel2}. $H(E)$ is the heel effect model, which can be simplified as a certain amount of the anode material's attenuation, and the amount of the attenuation mainly depends on the emitting angle\cite{Braun2010Heel}. $F(E)$ is the attenuation of regular filters in the CT system. $D(E)$ is the energy response of the detector.


\subsubsection{Filter Weights and Thicknesses Estimation}
Based on the multi-ray spectrum estimation model, we can model the effects of spectral filter by using the overdetermined nonlinear system of equations (\ref{eq:transmission_model}). In this work, we use the Levenberg-Marquardt (LM) method for least squares to solve for the parameters $w_{k}, T_{k}$. And multiple initial values can be preset to avoid non-convergence, such as $(w_1=1, w_2=0, T_1=T_{1,{\rm min}}, T_2=T_{2,{\rm min}})$, $(w_1=0.5, w_2=0.5, T_1=T_{1,{\rm max}}, T_2=T_{2,{\rm max}})$, where $T_{\rm min},T_{\rm max}$ can be obtained from known filter information. The LM method updates the parameter vector $x$ as:
\begin{eqnarray}
	\mathbf{x}_{l+1} &= \mathbf{x}_l - \left( \mathbf{J}_l^{\rm T}\mathbf{J}_l+\lambda\mathbf{I} \right)^{-1}\mathbf{J}_l^{\rm T}(\mathbf{A_{\rm measured}}-\mathbf{A}_l) \\
	\mathbf{x}_l &= (w_{1}, T_{1}, ... , w_{K}, T_{K})_l \\
	\mathbf{A}_l &= (A_{1}, A_2, ..., A_m, ... , A_{M})_l 
	\label{eq:LM}
\end{eqnarray}
where,
$\mathbf{x}_l$ is the parameter vector at $l_{\rm th}$ iteration; 
$\mathbf{r}_l$ is the residual vector at $l_{\rm th}$ iteration; 
$\lambda$ is the damping parameter;
$\mathbf{I}$ is the identity matrix;
$\mathbf{A}_{\rm measured}$ is the measured transmissions of the spectral filter using multiple altered initial spectra, and $\mathbf{A}_l$ is the transmission calculated by equation (\ref{eq:transmission_model}) with $\mathbf{x}_l$;
$\mathbf{J}_l$ is the Jacobian matrix of the residuals at $l_{\rm th}$ iteration. $\mathbf{J} \in \mathbb{R}^{M\times 2K}$ is,
\begin{eqnarray}
	\mathbf{J}(m,k) = \left\{
	\begin{array}{l l}
		& -\partial A_m / \partial w_k, \quad \mbox{if } k \mbox{ is odd,} \\
		& -\partial A_m / \partial L_k, \quad \mbox{if } k \mbox{ is even.}
	\end{array} \right.
\end{eqnarray}

\subsubsection{Modulated Spectrum Generation} 
Based on the weighted multi-ray spectrum estimation model, we can obtain the mixed spectra by equation (\ref{eq:spec_model_new}) with multiple estimated filter weights and thicknesses.

\subsubsection{Spectral Imaging} 
Finally, we can conduct beam hardening correction (BHC) or material decomposition for the polychromatic projections. The beam hardening artifacts and CT number inaccuracy can be well suppressed by the more accurate spectrum estimation.

\subsection{Validation Cases}
\begin{figure}[b]
	\centering
	\includegraphics[width=12cm]{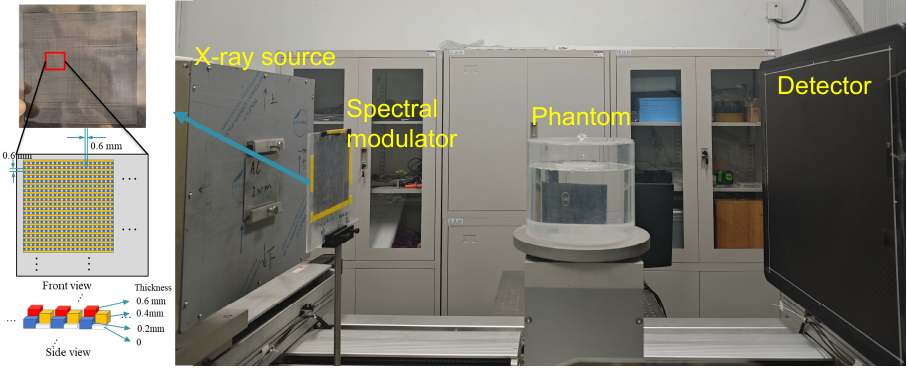}
	\caption{The tabletop CBCT system with a spectral modulator and a water phantom.} \label{fig:platform}	
\end{figure}

\begin{enumerate}
	\item Setup of spectral filters \quad
	In simulations, we used both a split filter and a molybdenum modulator as the spectral filters. In experiments, we used both a copper and a molybdenum modulator for testing. The split filter is a tin-gold (0.7 mm Sn, 0.07 mm Au) filter based on the second generation of split filter dual-energy CT\cite{Joël2023split}; the molybdenum modulator used in this work is chosen the same as the modulator we used before\cite{Deng2024Modulator}, which is formed by overlapping two 1D strip modulators, and has four kinds of filters (0, 0.2, 0.4, 0.6 mm thick); the copper modulator is a similar overlapped 2D modulators with the same thicknesses. Figure \ref{fig:platform} illustrates the structure of the modulators we used in experiments.
	
	\item Setup of numerical simulations \quad 
	In order to simulate the penumbra effects caused by the focal spot size, we set 3$\times$3 oversampling to the 0.5 mm focal spot; and 3$\times$3 oversampling to each detector pixel; clinical abdominal CT images obtained from Pancreas-CT Dataset\cite{roth2016data} were used to show the beam hardening artifacts caused by inaccurate spectral estimation. The main parameters of simulations are summarized in Table \ref{tab:parameters_all}. 
	It should be noted that in simulations, we conducted step experiments with 1, 2, 3, 4, 5, and 6 mm of Al filters and 0.3, 0.6, 0.9 mm Cu filters for the A-SWIFT method.
	
	\item Setup of physics experiments \quad
	In experiments, we scanned a pure water phantom and a multi-energy CT phantom (Gammex, Middleton, WI) with several iodine inserts on our tabletop CBCT system as shown in Figure \ref{fig:platform}. The x-ray source used a Varex G-242 tube with a focal spot of 0.4 mm; the detector was Varex 4343 RF flat-panel detector (FPD); the vertical collimation was reduced to 10 mm at the detector to reduce scatter signals; 
	for the water phantom, the X-ray source operated at 120 kVp just for beam hardening correction; for the Gammex phantom, the X-ray source operated at 80 and 120 kVp as a sequential scan for material decomposition.
	The blockers of the spectral modulator corresponding to the fan-beam scanning region is (0,0.4 mm Mo) and (0,0.4 mm Cu) for the Mo and Cu modulator, respectively. Similar step experiments with 1, 2, 3, 4, 5, and 6mm of Al filters and 0.3, 0.6, 0.9 mm Cu filters were also conducted. 
	
	\item Evaluations \quad
	In addition to evaluating the normalized root mean square error (NRMSE) between the estimated and reference spectra (mentioned in \ref{sec:analysis}), we further quantified the performance of the spectrum estimation by measuring the root mean square error (RMSE) of the mean values within the selected regions of interest (ROIs) in reconstructed images. The evalution metric $E_{\rm RMSE}$ is defined by,
	\begin{eqnarray}
		E_{\rm RMSE} = \sqrt{\frac{1}{N}\sum_{i=1}^{N}(\mu_i-\overline{\mu}_i)^2}
	\end{eqnarray}
	where $i$ is the index of the the ROIs,, N is the total number of the ROIs, $\mu_i$ is the averaged value in HU inside the $i_{\rm th}$ ROI, and  $\overline{\mu}_i$ is the corresponding value on the reference image.
\end{enumerate}

\begin{table}[h]
	\centering
	\caption[]{\upshape Main parameters in numerical simulations and physics experiments. SID: source-to-isocenter distance; SDD: source-to-detector distance.} \label{tab:parameters_all}
	{
		\begin{tabular}{cccc}
			\hline\hline
			\rule{0pt}{8pt}
			Parameters & Simulations- & Simulations-
			& Experiments\\  
			& split filter & modulator& \\
			\hline
			\rule{0pt}{8pt}
			Focal spot size 	& 0.5 mm & 0.5 mm & 0.4 mm \\
			Detector pixel size & 0.6 mm & 0.3 mm & 0.3 mm \\
			Detector matrix size& $800\times8$ & $1440\times 1$ & $1440\times1440$\\
			Views per rotation 	& 500 & 720 & 720 \\
			Number of rotations	& 10   & 1 &1  \\
			Scanning geometry & helical & circular & circular \\
			Pitch & 0.5 & 0 &  0 \\
			SID & 800 mm  & 800 mm  & 750 mm \\
			SDD & 1200 mm & 1200 mm & 1180 mm \\
			Tube volatage & 120 kVp & 120 kVp & 80 / 120 kVp\\
			\hline\hline
		\end{tabular}
	}
\end{table}

\section{Result}
\subsection{Numerical Simulations}
\begin{figure}[tb]
	\centering
	\includegraphics[width=16cm]{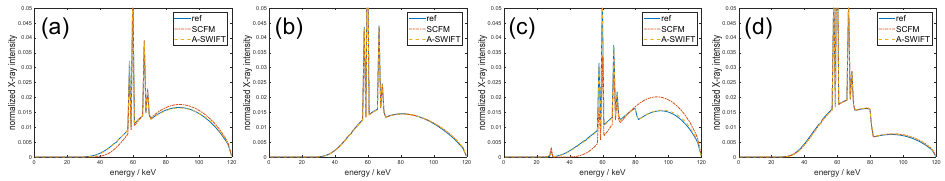}
	\caption{Spectra calibrated by SCFM and A-SWIFT methods in different regions. (a): penumbra region of spectral modulator (boundary between 0.2 mm and 0.6 mm Mo blockers); (b): umbra area (0.2 mm Mo); (c): penumbra region of split filter (boundary between 0.07 mm Au and 0.7 mm Sn filters); (d): umbra area (0.07 mm Au).} \label{fig:simulation_spectra}	
\end{figure}

Figure \ref{fig:simulation_spectra} presents the estimated spectra for both the spectral modulator and split filter, obtained using the A-SWIFT method and SCFM method.
As shown in Figure \ref{fig:simulation_spectra} (a)(c), in the penumbra region, the spectra estimated by A-SWIFT are much closer to the true mixed spectrum compared to those from SCFM. In the umbra area (shown in Figure \ref{fig:simulation_spectra} (b)(d)), both methods accurately fit the true spectrum, demonstrating the robustness of the A-SWIFT method. 

Table \ref{tab:NRMSE} displays the mean energy bias and NRMSE corresponding to the spectra in Figure \ref{fig:simulation_spectra}. The mean energy bias is defined as the mean energy of the estimated spectra minus the mean energy of the simulated spectra, while NRMSE is widely used for assessing spectrum estimation accuracy\cite{ChangIEEE2024}. The decrease in these metrics demonstrates the effectiveness of the A-SWIFT method for spectral mixing.

\begin{table}[ht]
	\centering
	\caption[]{\upshape Quantitative analysis of estimated spectra by A-SWIFT and SCFM in the penumbra region (Figure \ref{fig:simulation_spectra} (a)(c) ).} \label{tab:NRMSE}
	{
		\begin{tabular}{ccc}
			\hline\hline
			\rule{0pt}{8pt}
			Spectrum & Mean energy bias (keV) & NRMSE(\%)\\  
			\hline
			\rule{0pt}{8pt}
			modulator SCFM  			& 1.98	& 6.5	\\ 
			modulator A-SWIFT  			& 0.61  & 2.0	\\ 
			split filter SCFM			& 7.43  & 32.7	\\ 
			split filter A-SWIFT		& 0.72  & 2.8	\\ 
			\hline\hline
		\end{tabular}
	}
\end{table}

\begin{figure}[tb]
	\centering
	\includegraphics[width=16cm]{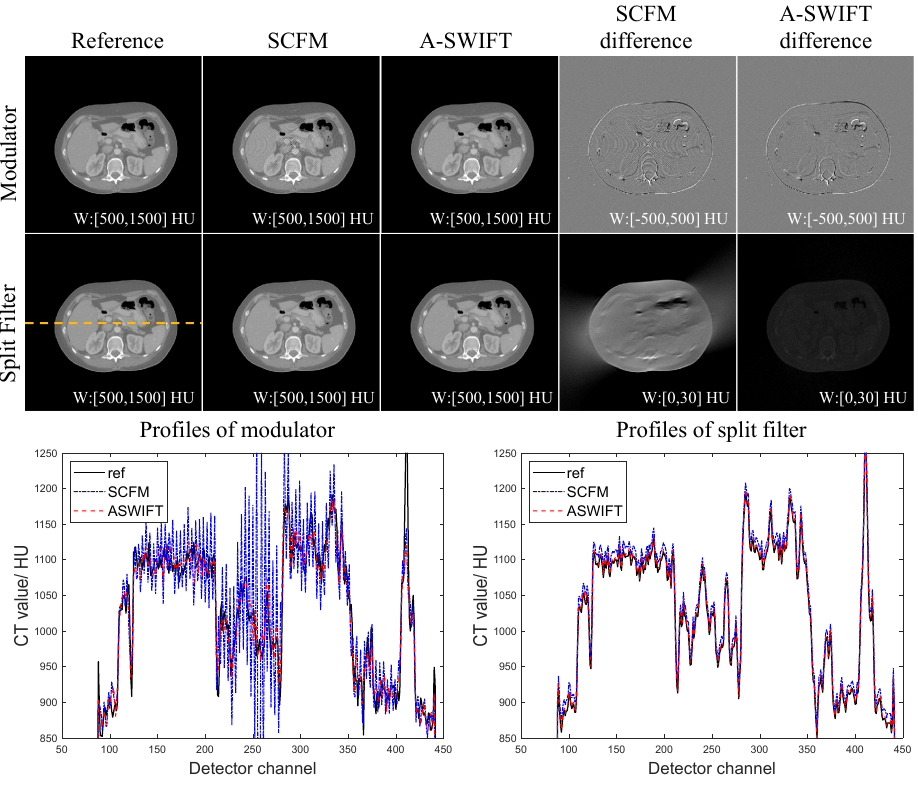}
	\caption{The beam hardening correction results by the SCFM and A-SWIFT method with both a Mo modulator and a tin-gold split filter. } \label{fig:simulation_result}	
\end{figure}

Figure \ref{fig:simulation_result} shows the beam hardening correction (BHC) results for both a Mo modulator (with 0.2 mm and 0.6 mm Mo blockers) and a split filter (0.7 mm Sn and 0.07 mm Au) with a water phantom. 

For the spectral modulator, the periodic arrangement of different submodules results in a periodic distribution of the penumbra region at the detector. Consequently, inaccurate spectrum estimation in the penumbra region leads to ring artifacts in the reconstructed images. The top row of Figure \ref{fig:simulation_result} displays the reconstructed images and the difference images for the modulator.  
The ring artifacts in the images are well suppressed in A-SWIFT results due to accurate spectrum estimation. 

For the split filter, which contains two different submodules (Sn, Au), the penumbra region only exists at the center row of the detector. Consequently, inaccurate spectrum estimation in the penumbra region results in slight artifacts and incorrect CT value in the reconstructed images. The bottom row of Figure \ref{fig:simulation_result} presents the reconstructed images and difference images for the split filter.  
The CT value inaccuracy can be effectively reduced by the A-SWIFT method, with the RMSE of the entire image decreasing from 8.6 HU for SCFM to 1.9 HU for A-SWIFT.

\subsection{Physics Experiments}
\begin{figure}[h]
	\centering
	\includegraphics[width=16cm]{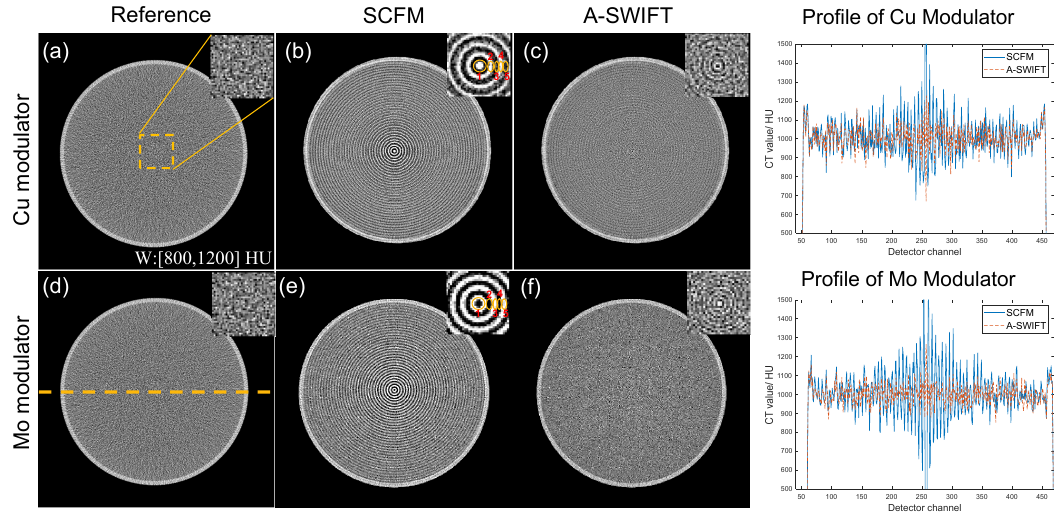}
	\caption{BHC results from physics experiments with a Cu or Mo modulator placed between the source and the water phantom using SCFM and A-SWIFT method for the spectrum estimation. Gray window: [800,1200] HU. The five ROIs labeled on the zoomed-in image in (b) are used for the analysis in Table \ref{tab:experiment_water}.} \label{fig:experiment_Mo_Cu_water}	
\end{figure}

Figure \ref{fig:experiment_Mo_Cu_water} shows the reconstructed images and profiles after BHC of the water phantom using both the molybdenum and copper modulator. The fan-beam CT image of the water phantom with no spectral modulator is shown in Figure \ref{fig:experiment_Mo_Cu_water} (a),(d), and taken as our reference image, where the ROIs are marked by yellow circle in Figure \ref{fig:experiment_Mo_Cu_water} (b). The ring artifacts caused by inaccurate spectrum estimation could be well suppressed by A-SWIFT method. Some residual artifacts in the (c),(f) zoom-in image could be due to the residual scatter.

Table \ref{tab:experiment_water} shows the mean value and standard deviation (STD) of the selected ROIs in Figure \ref{fig:experiment_Mo_Cu_water}. The A-SWIFT method shows much better performance in CT number accuracy, with the $E_{\rm RMSE}$ of all ROIs reduced from 85 HU to 21 HU for the copper modulator, and from 77 HU to 7 HU for the molybdenum modulator.
Even for spectral modulators with extensive penumbra regions and severe spectral mixing effects, the A-SWIFT method still performs well for most pixels in penumbra region.

\begin{table}[tb]
	\centering
	\caption[]{\upshape Mean and standard deviation (STD) of the selected ROIs (marked by the yellow circle) in Figure \ref{fig:experiment_Mo_MEPhan}. (The reference mean for all ROIs is 1000 HU)} \label{tab:experiment_water}
	{
		\begin{tabular}{cccccccccc}
			\hline\hline
			\rule{0pt}{8pt}
			 & Modulator & Method  & ROI 1 & ROI 2 & ROI 3 & ROI 4 & ROI 5& $E_{\rm RMSE}$ & Avg.\\
			\hline
			\rule{0pt}{8pt}  \multirow{4}{*}{\shortstack{MEAN \\ (HU)}} &
			\multicolumn{2}{c}{No-Modulator} & 
			994 & 987 & 1006 & 1005 & 968 & 16 & /
			\\
			\cline{2-10} &
			\multirow{2}{*}{Cu}
			  & SCFM 	  & 1029 & 924 & 1067 & 861 & 1078 & 85 & /	\\
			  & & A-SWIFT  & 1007 & 973 & 987 & 981 & 1028 & 21 & /	\\
			 \cline{2-10} &
			\multirow{2}{*}{Mo}
			  & SCFM 	 & 1100 & 986 & 1038 & 887 & 1071 & 77	 & /\\
			  & & A-SWIFT  & 1011 & 1004 & 1003 & 993 & 995  & 7  & /\\
			  \hline
			  \rule{0pt}{8pt}  \multirow{5}{*}{\shortstack{STD \\ (HU)}} &
			  \multicolumn{2}{c}{No-Modulator} & 
			  68 & 69 & 60 & 64 & 55 & / & 63
			  \\
			  \cline{2-10} &
			  \multirow{2}{*}{Cu}
			  & SCFM 	  & 361 & 201 & 196 & 91 & 160 & / & 202	\\
			  & & A-SWIFT  & 140 & 102 & 108 & 91 & 58 & / & 100	\\
			  \cline{2-10} &
			  \multirow{2}{*}{Mo}
			  & SCFM 	 & 471 & 293 & 258 & 218 & 191 & /	 & 286\\
			  & & A-SWIFT  & 103 & 103 & 79 & 100 & 97  & /  & 96\\
			 \hline\hline
		\end{tabular}
	}
\end{table}

\begin{figure}[h]
	\centering
	\includegraphics[width=15cm]{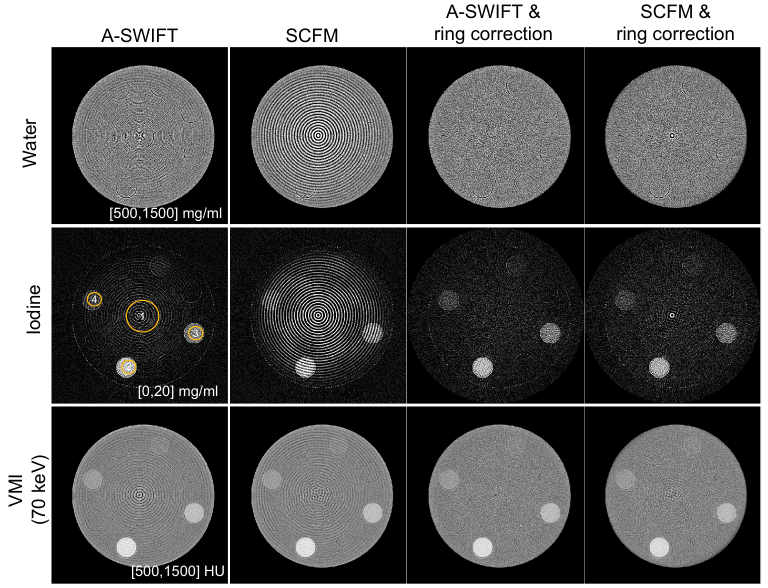}
	\caption{Material decomposition results from physics experiments with a Mo modulator and the Gammex phantom using SCFM and A-SWIFT method for the spectrum estimation.} \label{fig:experiment_Mo_MEPhan}	
\end{figure}

Furthermore, Figure \ref{fig:experiment_Mo_MEPhan} presents the basis material images and virtual monochromatic images (VMIs) of the Gammex phantom using the molybdenum modulator, and the material decomposition was conducted using a polynomial fitting method. The residual ring artifacts observed in the images produced by the A-SWIFT method could be attributed to the deviation of the initial spectrum without modulator, as well as residual scatter and noise, since the material decomposition is highly sensitive to these factors. To provide a better comparison between the A-SWIFT method and the SCFM method, a ring correction method\cite{Wang2024Ring} was also applied. The ring artifacts are effectively suppressed by A-SWIFT when combined with the ring correction, whereas the ring correction struggles to address the severe ring artifacts produced by the inaccurate spectrum estimation of the SCFM method and changes the mean value of selected ROIs.

Table \ref{tab:experiment_Mo_MEPhan} provides the mean and STD of ROIs shown in Figure \ref{fig:experiment_Mo_MEPhan}. The $E_{\rm RMSE}$ of all ROIs in iodine images is effectively reduced by A-SWIFT compared to SCFM, with the mean values obtained using A-SWIFT with or without ring correction are both closer to the ground truth than those obtained using SCFM.
Besides, the STD in all ROIs of A-SWIFT methods without ring correction shows better performance than SCFM method, and is close to the STD obtained by both methods combined with ring correction, which can be regarded as a references that only caused by noise, as ring artifacts are almost eliminated in the iodine images. These quantitative analysis demonstrates the improved accuracy of spectrum estimation by A-SWIFT method.


\begin{table}[h]
	\centering
	\caption[]{\upshape Mean and standard deviation (STD) of estimated material densities (ROI 1-4) in Figure \ref{fig:experiment_Mo_MEPhan}. The results of A-SWIFT method with ring correction are regarded as the  reference for STD. (RC: ring correction; ref: reference)} \label{tab:experiment_Mo_MEPhan}
	{
		\begin{tabular}{cccccccc}
			\hline\hline
			\rule{0pt}{8pt}
			 &  Method & \shortstack{1000\\ mg/ml\\ Water} & \multicolumn{1}{|c}{\shortstack{15.0 \\mg/ml\\ Iodine}} & \shortstack{10.0\\ mg/ml\\ Iodine} & \shortstack{5.0\\ mg/ml\\ Iodine}  & \shortstack{ Iodine \\ $E_{RMSE}$} & \shortstack{ Iodine \\ Avg.}\\
			\hline
			\rule{0pt}{8pt} 
			\multirow{4}{*}{Mean}
			& SCFM 	  & 992 & \multicolumn{1}{|c}{14.2}  & 9.3  & 4.9 & 0.6 & /	\\
			& A-SWIFT & 991 & \multicolumn{1}{|c}{14.9}  & 9.7  & 4.8 & 0.2 & /		\\
			\cline{2-8}
			& SCFM \& RC    & 987 & \multicolumn{1}{|c}{13.5} & 8.7 & 4.5  & 1.2 & /	\\
			& A-SWIFT \& RC & 990 & \multicolumn{1}{|c}{14.5} & 9.3 & 4.5 & 0.6 & /	\\
			\hline
			\multirow{4}{*}{STD}
			& SCFM 	 & 539 & \multicolumn{1}{|c}{9.9} & 8.9	& 10.4 & / & 9.7	\\
			& A-SWIFT & 312 & \multicolumn{1}{|c}{6.3} & 5.9	& 5.5 & / & 5.9	\\
			\cline{2-8}
			& SCFM \& RC  	& 257 & \multicolumn{1}{|c}{5.7} & 5.6 & 4.9 & / & 5.4	\\
			& A-SWIFT \& RC  & 234 & \multicolumn{1}{|c}{5.8} & 5.6 & 4.8 & / & 5.4		\\
			\hline\hline
		\end{tabular}
	}
\end{table}

\section{Discussions and Conclusions}
\label{sec:Conclusions}
Traditionally, modeling the focal spot of an X-ray source has been a key approach to mitigating image blurring caused by penumbra effects. However, for penumbra-effect induced spectral mixing, focal spot modeling alone is insufficient to restore the mixed spectrum accurately, nor does it ensure that the restored spectrum matches the corresponding deblurred projections. To address these challenges, we propose a multi-ray model combined with the A-SWIFT method to approximate the spectra affected by spectral mixing in the penumbra region.
Simulation and experimental results demonstrate that this method enhances spectral estimation accuracy and suppresses image artifacts caused by inaccurate spectrum estimation. However, our method also has certain limitations.
\begin{enumerate}
	\item The multi-ray model remains an approximation of spectral mixing caused by X-rays from the focal spot. Due to the approximation, the model maintains systematic bias, as accurately representing the effects of infinite spectral mixing is unattainable. In simulations, this limitation results in slight inherent errors, which can be observed as residual mean energy bias in the estimated spectrum.
	Fortunately, in practical experiments, these slight errors are often masked by noise, and increasing the number of weights and filters in the multi-ray model could reduce this bias. 
	\item The inaccuracy of initial spectrum estimation of the CT system (excluding the spectral filter) affects the performance of A-SWIFT method. Although some measurement methods like EM method can be used to reduce these deviations, the initial spectrum deviation still exists in reality, and impacts the spectral imaging results, especially the material decomposition, which is very sensitive to spectral accuracy. As shown in Figure \ref{fig:experiment_Mo_MEPhan}, initial spectral estimation errors are likely the another reason of the residual ring artifacts besides the residual scatter. 
	\item The residual ring artifacts remains in the spectral imaging results. Although the ring artifacts could be effectively suppressed by our method, the residual ring artifacts still affects the image quality. Fortunately, the ring artifacts is slight and can be easily removed by some ring removal methods.
\end{enumerate}

Overall, this method maintains compatibility with existing methods for regular filters or regions without spectral mixing, and shows great advantages for spectral filters in the penumbra regions with spectral mixing.

\section*{Acknowledgments}
This project was supported in part by grants from the National Natural Science Foundation of China (No. 12075130 and No. U20A20169), and in part by the National Key R\&D Program of China (No. 2022YFE0131100).

\section*{Data Availability Statement}
The data that support the findings of this study are available upon reasonable request from the authors.
\newcommand{\newblock}{}
\bibliographystyle{medphy}
\bibliography{ref}
\end{document}